\documentclass[11pt]{article}
\usepackage{amssymb}
\usepackage{epsfig,color}
\usepackage{pstricks,graphicx,epsfig,color,amssymb,amsmath,amscd}

\usepackage{cite}
\usepackage{latexsym}
\usepackage[]{graphicx}
\usepackage{makeidx}
\usepackage[]{caption}

\newcommand{\be}{\begin{eqnarray}}
\newcommand{\ee}{\end{eqnarray}}

\begin{document}

\title{Superheavy dark matter in $R^2$-cosmology }

\begin{titlepage}
\author{E.~V.  Arbuzova}

\maketitle
\begin{center}
Department of Higher Mathematics, Dubna State University,  Universitetskaya str.~19, Dubna, 141983, Russia \\
Department of Physics, Novosibirsk State University, Pirogova str.~2, Novosibirsk, 630090, Russia
\end{center}

\begin{abstract}
The conventional Friedmann cosmology is known to be in tension with the existence of stable particles having interaction strength typical for supersymmetry and heavier than several TeV. A possible way to save life of such particles may be a modification of the standard cosmological expansion law in such a way that the density of these heavy relics would be significantly reduced. We study particle creation in the Starobinsky inflationary model for different decay channels of the scalaron. It is shown that in the process of thermalization superheavy stable particles with the coupling strength typical for the GUT SUSY could be created with the density equal to the observed density of dark matter.
\end{abstract}

\thispagestyle{empty}
\end{titlepage}

\section{Introduction}

An accepted property of dark matter (DM) particles is that they are electrically neutral, since they do not scatter light, hence named {\it Dark } (matter). Otherwise their properties are practically unknown. This opens possibilities for particles of many different types to be dark matter candidates.  

A natural and, formerly, very popular candidate for dark matter particle is the lightest supersymmetric particle (LSP), which should be stable, if the so-called R-parity is conserved. However, an extensive search for the low energy supersymmetry at LHC led to negative results. Thus, if supersymmetry exists, its characteristic energy scale should be higher than, roughly speaking, 10 TeV. Though no direct limits from below on LSP mass were presented, a study of higher energy SUSY and heavier LSPs may be interesting.    

To estimate the energy density of LSPs one can use the ratio of the contemporary number density of $X$-particles, $n_X$, to the number density of CMB photons, 
$n_{\gamma}$:
\be
\frac {n_X}{n_\gamma} \approx \frac{1}{m_{Pl} m_X \sigma_{ann} v}, 
\label{n-X-to-n-gamma} 
\ee
where $m_{Pl} = 1,2 \cdot 10^{19} $ GeV is the Planck mass. 
This important result was first obtained by Zeldovich~\cite{Zeldovich:1965gev} and later rederived in detail in several textbooks~\cite{Kolb:1990vq,Gorbunov:2011zz}.  

For S-wave annihilation the product of the annihilation cross-section of $X\bar X$-particles with mass $m_X$ by their center-of-mass velocity is
by the order of magnitude:
\be 
\sigma_{ann} v  = \frac{\alpha^2} { m_X^2},
\label{sigma-v}
\ee
where $\alpha $ is a coupling constant. 
With the typical for SUSY coupling, $\alpha \approx 0.01$, the energy density of LSPs can be estimated as
\be
\rho_{LSP} \sim \rho_{DM}^{(obs)} (M_{LSP}/ 1\,\rm{TeV})^2 ,
\label{rho-LSP}
\ee
where  $ M_{LSP} $ is the mass of LSP and $\rho_{DM}^{(obs)} \approx 1$ keV/cm$^3$ is the observed 
value of the cosmological density of dark matter.

The cosmological energy density of LSPs is proportional to their mass squared and for the LSPs with the the mass of order 1 TeV, their energy density is of the order 
of the observed dark matter energy density. For larger masses LSPs would overclose the universe. 
These unfortunate circumstances exclude LSPs as dark matter particles in the conventional cosmology. However, in $R+R^2$ gravity the energy density 
of LSPs may be significantly reduced, so it reopens for them the chance to be viable candidates for dark matter, if their mass is larger than 1000 TeV.

\section{Cosmological evolution in $R^2$ modified gravity}

The theory of gravitational interactions, General Relativity (GR), based on the Einstein-Hilbert action 
\be
S_{EH} = -\frac{m_{Pl}^2}{16\pi} \int d^4 x \sqrt{-g}\, R,  
\label{act-EH}
\ee
describes basic properties of the universe in very good agreement with observations. However, some features of the universe may request to go beyond the frameworks of GR. In theories of $F(R)$-modified gravity it is achieved by an addition of a non linear function of curvature, $F(R)$, into the usual GR action (\ref{act-EH}). 

In 1979 V.T. Gurovich and A.A. Starobinsky suggested to take $F(R)$ proportional to the curvature squared for the elimination of cosmological singularity 
\cite{Gurovich:1979xg}. In the subsequent paper by Starobinsky \cite{Starobinsky:1980te} it was found that  
the addition of the $R^2$-term leads to inflationary cosmology. In what follows we consider the theory with the action: 
\be
S_{tot} = -\frac{m_{Pl}^2}{16\pi} \int d^4 x \sqrt{-g} \left(R-\frac{R^2}{6m^2_{R}}\right)+S_m\,,
\label{S-R2-tot-1}
\ee
where 
$S_m$ is a matter action. Curvature ${ R(t)}$ can be considered as an effective scalar field (scalaron) with the mass $ m_R$ and with the decay width 
$ \Gamma $.

According to calculations presented in Ref.~\cite{Arbuzova:2018ydn} the cosmological evolution in $R^2$-theory can be separated into four distinct epochs. 

Firstly, there was the inflationary stage, when the curvature was sufficiently large and the universe expanded exponentially with slowly decreasing $R(t)$. 

The second epoch began when $R(t)$ approached zero and started to oscillate around it, periodically changing sign:
\be 
R=-\frac{4m_R\cos(m_Rt+ \theta)}{t}. \label{rsol}\, 
\ee
At this stage the Hubble parameter oscillates almost touching zero:
\be 
H\equiv \frac{\dot a}{a} = \frac{2}{3t} \left[1+ \sin(m_Rt + \theta )\right]  \label{hsol}.
\ee
Energy density of matter at this stage drops down as 
\be 
\rho_{R^2} = \frac{m^3_R}{120 \pi t},
\label{rho-R2}
\ee
 instead of GR behaviour 
\be
\rho_{GR} = \frac{3H^2m_{Pl}^2}{8 \pi} = \frac{3 m_{Pl}^2}{ 32 \pi t^2}.
\label{rho-GR}
\ee
The oscillations of curvature led to particle production and this epoch can be called Big Bang. 

This stage was followed by the  transition period from the scalaron domination to domination of the usual (relativistic) matter. Finally, after the complete decay of scalaron we arrived to the conventional cosmology governed by General Relativity.   

The cosmological energy density of matter at scalaron domnated period depends upon the decay width of the scalaron, which, in turn, depends
upon the dominant decay channel. 

If there exists scalar particle minimally coupled to gravity, the decay width of the scalaron into massless scalars does not vanish in the limit of zero 
mass and would be: 
\be
\Gamma_S = \frac{m_R^3}{24 m_{Pl}^2}.
\label{Gamma-S}
\ee
 In this case, the energy density of predominantly relativistic matter is  equal to:
\be 
\rho_{S} (t) = \frac{m_R^3}{120\pi t}. 
\label{rhos-of-t}
\ee
If scalaron predominantly decays into a pair of fermions or conformally coupled scalars with the mass $m_f$, the decay width is proportional to the particle mass squared and 
the decay probability is dominated by the heaviest final particle:
\be
\Gamma_f = \frac{ m_R m_f^2 }{24 m_{Pl}^2 }. 
\label{Gamma-f}
\ee
The corresponding energy density of matter is:
\be 
\rho_{f} (t) = \frac{ m_R m_f^2}{120 \pi t}.
\label{rhof-of-t}
\ee
Finally, if scalaron decay is induced by the conformal anomaly, the decay width of scalaron into gauge bosons would be equal to:
\be
\Gamma_{an} = \frac{\beta_1^2 \alpha^2 N}{96\pi^2}\,\frac{m_R^3}{m_{Pl}^2} ,
\label{Gamma-an}
\ee
where $\beta_1$ is the first coefficient of the beta-function, $N$ is the rank of the gauge group, and
$ \alpha$ is the gauge coupling constant, which at high energies depends upon the theory. 
Correspondingly, the energy density of matter created by the decay into this channel would be:
\begin{equation}
\rho_{an} = \frac{\beta^2_1 \alpha^2 N}{4 \pi^2} \,\frac{m_R^3}{120 \pi t}. 
\label{rho-an-of-t}
\end{equation}

The presented laws (\ref{rhos-of-t}), (\ref{rhof-of-t}), and (\ref{rho-an-of-t}) demonstrate much slower decrease of the energy density of matter than 
usually for relativistic matter at scalaron dominated epoch, where $\rho \sim 1/a^4(t) \sim 1/t^{8/3}$, since the scale factor at SD stage $a(t) \sim t^{2/3}$.
It is ensured by the influx of energy from the scalaron decay. 

It is interesting to compare equations connecting temperature with time for different expressions for energy density of matter. In thermalized plasma 
the energy density of relativistic matter is proportional to the forth power of temperature:
\be 
\rho_{therm} = \frac{\pi^2 g_* T^4}{30},
\label{rho-pl}
\ee
where $g_*$ is the number of relativistic species in the plasma,  $g_* \sim 100$.

Equating critical energy density of matter in GR (\ref{rho-GR}), as well as energy densities of matter (\ref{rhos-of-t}), (\ref{rhof-of-t}), and (\ref{rho-an-of-t})
in $R^2$-theory, to the energy density of relativistic plasma in thermal equilibrium, we obtain  that the connection of the temperature with time has 
very different forms  in General relativity and in $R^2$-cosmology:
\be \label{tT-GR}
(t T^2 )_{GR} &=& \left(\frac{ 90}{32 \pi^3 g_*}\right)^{1/2} \,m_{Pl} = const; \\ 
(t T^4)_s &=& \frac{ m_R^3}{4 \pi^3 g_*} = const;\\
(t T^4)_f &=& \frac{m_R m_f^2}{4\pi^3 g_*} = const;\\ \label{tT-an} 
(t T^4)_{an} &=& \frac{0.78 }{\pi^2  g_*}\,  \alpha_R^2 m_R^3 = const.
\ee
We see that the canonical relation between the temperature of matter and the cosmological time in General Relativity, $T^2 t = C_{GR} $, 
is replaced in $R^2$-theory by the relation $T^4 t =  C_{R^2}$. Moreover, as it is followed from  (\ref{tT-GR}),  $C_{GR}$ is a universal constant, proportional to the Planck mass, while in $R^2$-cosmology the constant $C_{R^2}$ depends on the model and may be strongly different for the scalaron decay into non-conformal massless bosons, fermions or gauge bosons. Note, that in Eq.~(\ref{tT-an}) the coupling ${\alpha_R}$ is taken at the energies equal to the scalaron mass.

\section{Evolution of $X$-particles in thermal plasma}

The freezing of massive species $X$ with  mass $M_X$ is governed by the following equation: 
\be  {
\dot n_X + 3H n_X = -\langle \sigma_{ann} v \rangle \left( n_X^2 - n^2_{eq} \right)}, \   
n_{eq} = g_s \left(\frac{M_X T}{2\pi}\right)^{3/2} e^{-M_X/T}, 
\label{dot-n-X}
\ee
where $n_X$ is the number density of $X$-particles, $n_{eq}$ is their equilibrium number density, $g_s$ is the number of spin states.
$ \langle \sigma_{ann} v \rangle$ is the thermally averaged annihilation cross-section of X-particles with $v$ being the center-of-mass velocity.

Equation (\ref{dot-n-X}) was derived in 1965 by Zeldovich \cite{Zeldovich:1965gev} and collaborators \cite{Zeldovic:1965rys,Zeldovic:1965UFN}. 
In 1977 it was applied to freezing of massive stable neutrinos \cite{Lee:1977ua,Vysotsky:1977pe} and after that this equation was named as 
the Lee-Weinberg equation, though justly it should be called the Zeldovich equation. 

For annihillation of non-relativistic particles in S-wave the thermal averaging over medium is not essential and the annihilation cross-section
can be estimated as:
\be 
\langle \sigma_{ann} v \rangle= \sigma_{ann} v = \frac{\alpha^2 \beta_{ann}}{M_X^2}. 
\label{sigma-ann-S}
\ee 
Since annihilation of Majorana fermions proceeds in P-wave, the thermal averaged annihilation cross-section acquires the factor $T/M_X$:
\be 
\langle \sigma_{ann} v \rangle =  \frac{\pi \alpha^2 \beta_{ann}}{M_X^2} \,\frac{T}{M_X}. 
\label{sigma-ann-P}
\ee
Here ${\beta_{ann}}$ is a numerical parameter 
proportional to the number of annihilation channels, ${\beta \sim 10}$. 

An important comment is of order here. There are two possible channels to produce massive stable $X$-particles: first, directly through the 
scalaron decay into a $X\bar X$-pair, and, second, by the inverse annihilation of relativistic particles in thermal plasma. 
Direct production of $X\bar X$-pair by scalaron leads to the conclusion that the energy density of $X$-particles in the present day universe would be equal to the 
observed energy density of dark matter 
\be
{\rho_X^{(0)} \approx \rho_{DM}}\approx 1 {\rm keV/cm}^3, 
\ee
if ${M_X \approx 10^7}$ GeV. On the other hand, for such a small mass thermal $X$-particle production (through inverse annihilation) would be too
strong and would result in very large density of $X$-particles. But for larger masses $\rho_X^{(0)}$ would be unacceptably larger than DM energy density. 

A possible way out from this "catch-22" is to find a mechanism to suppress the scalaron decay into a pair of $X$-particles. And such mechanism does exist.  Since oscillating curvature scalar creates particles only in a symmetric state, 
{the direct production of $ X$-particles is forbidden, if they are Majorana fermions}, which must be in an antisymmetric state.   

Firstly, let us consider the scalaron decay into massless non-conformal scalars (detailed calculations are presented in our paper \cite{Arbuzova:2018apk}) . The dimensionless Zeldovich equation (\ref{dot-n-X}) has the form:
\be
\frac{df}{dx} =  {  - \frac{0.12 g_s \alpha^2 \beta_{ann}}{\pi^3 g_*} \left(\frac{m_R}{M_X}\right)^3} \, \frac{f^2 - f_{eq}^2}{x^5},
\label{Zeq-s}
\ee
where $x=M_X/T$ is a dimensionless new variable and the new dimensionless function $f$ is introduced according to
\be
n_X = n_{in} \left(\frac{a_{in}}{a}\right)^3, 
\ee
where  $n_{in}$ is the value of $X$-particle density at $a=a_{in}$ and $T_{in}=M_X$, so the $X$-particles can be considered as relativistic and
thus 
\be
n_{in} = 0.12 g_s T^3_{in} = 0.12 g_s M_X^3. 
\label{n-in}
\ee

Taking the following values of parameters: $g_* =100$, $\alpha = 0.01$, $\beta_{ann} =10$, and $m_R = 3\times 10^{13} $ GeV, 
we estimate the present day energy density of $X$-particles as:
\be 
\rho_X  \approx  1.7\times 10^8 \left(\frac{10^{10} {\rm {GeV} }}{M_X} \right)} \,{ \rm{keV/cm^3}.
\label{rho-X}
\ee
 This value is to be compared with the observed energy density of dark matter, ${ \rho_{DM} \approx 1}$ keV/cm$^3$. We see that 
$X$-particles must have  huge 
mass much higher than the scalaron mass, 
$M_X \gg m_R$,  to make reasonable DM density. However, if ${M_X > m_R}$, then classical scalaron field can still create $X$-particles, but the probability of their production would be strongly suppressed and such LSP with the
mass somewhat  larger than ${m_R}$ could successfully make the cosmological dark matter.

As the next step, let us turn to the scalaron decay into fermions or massive conformal scalars (for details see \cite{Arbuzova:2018apk}). If  bosons are conformally invariant due to non-minimal coupling to curvature, as $\xi R \phi^2 $ with $\xi = 1/6$, they are not produced if their mass is zero. The probability of production of both bosons and fermions is proportional to their mass squared.  In what follows we confine ourselves to consideration of fermions only. The width and the energy density of the scalaron decay into a pair of fermions are given by  expressions (\ref{Gamma-f}) and (\ref{rhof-of-t}) and the largest contribution into the cosmological energy density at scalaron dominated regime is presented by the  decay into the heaviest fermion species. 

We assume, that the mass of the LSP is considerably smaller than the masses of the other decay products, ${m_X < m_f}$, at least as 
${m_X \leq 0.1 m_f}$. In this case the direct production of $X$-particles by ${R(t)}$  can  be neglected and LSPs are dominantly produced by the secondary reactions in the plasma, which was  created by the scalaron decay into heavier particles.

Dimensionless kinetic equation for freezing of fermionic species takes the form:
\be 
\frac{df}{dx} =  {- \frac{\alpha ^2 \beta_{ann}}{\pi ^3 g_*} \, \frac{ n_{in}\,m_R m_f^2}{m_X^6} }\,  \, \frac{f^2 - f_{eq}^2}{x^5}, 
\label{df-dx-R2-2-f}
\ee 
where $n_{in}=0.09 g_s m_X^3$ is the initial number density of $X$-particles at  $T \sim m_X$. 

The contemporary energy density of $X$-particles can be approximately estimated as 
\be
\rho_X =  7 \cdot 10^{-9} \frac{m_f^3}{m_X m_R}\, {\rm cm^{-3}},
\label{rho-fr-fn} 
\ee
where we have taken $\alpha = 0.01$, $\beta_{ann} = 10$, $g_*=100$.

If in Eq.~(\ref{rho-fr-fn}) we put ${m_f = 10^5}$ GeV and ${m_X = 10^4}$~GeV, then  the energy density of $X$-particles will be much less than the 
observed dark matter energy density: $\rho_X \ll \rho_{DM}$. However, choosing ${m_X \sim 10^6}$ GeV, ${m_f \sim 10^7}$ GeV we obtain that 
${\rho_X}$ becomes comparable with 
the energy density of the cosmological DM, ${\rho_{DM}~\approx~1}$ keV/cm$^3$:
\be 
\rho_ X = 0.23\,  \left(\frac{m_f}{10^7\, {\rm GeV}}\right)^3 \left(\frac{10^6\, {\rm GeV}}{m_X}\right) \, {\rm \frac{keV}{cm^3}}.
\label{rho-fin}
\ee

Finally, let us consider the third possibility to produce $X$-particles through the scalaron decay into gauge bosons induced by the conformal anomaly 
\cite{Arbuzova:2020etv}. We assume that $X, \bar X$ are Majorana fermions and the direct production of them  by scalaron is forbidden. In this 
case $X\bar X$-pairs are produced through the inverse annihilation of relativistic particles in the
thermal plasma. 

To find the energy density of the produced $X$-particles in this case we solve again the Zeldovich equation (\ref{dot-n-X}) analytically and numerically.  
In the Fig.~\ref{rho-of-MX} the logarithm of the energy density of $X$-particles with respect to the observed energy density of dark matter as a function of the $X$-particle mass, $M_X$ is presented. 
It is clearly seen that 
$X$-particles may be viable candidates for the 
carriers of the cosmological dark matter, if their mass ${M_X \approx 5\cdot 10^{12}}$ GeV.  

\begin{figure}[!htbp]
  \centering
 \includegraphics[scale=0.5]{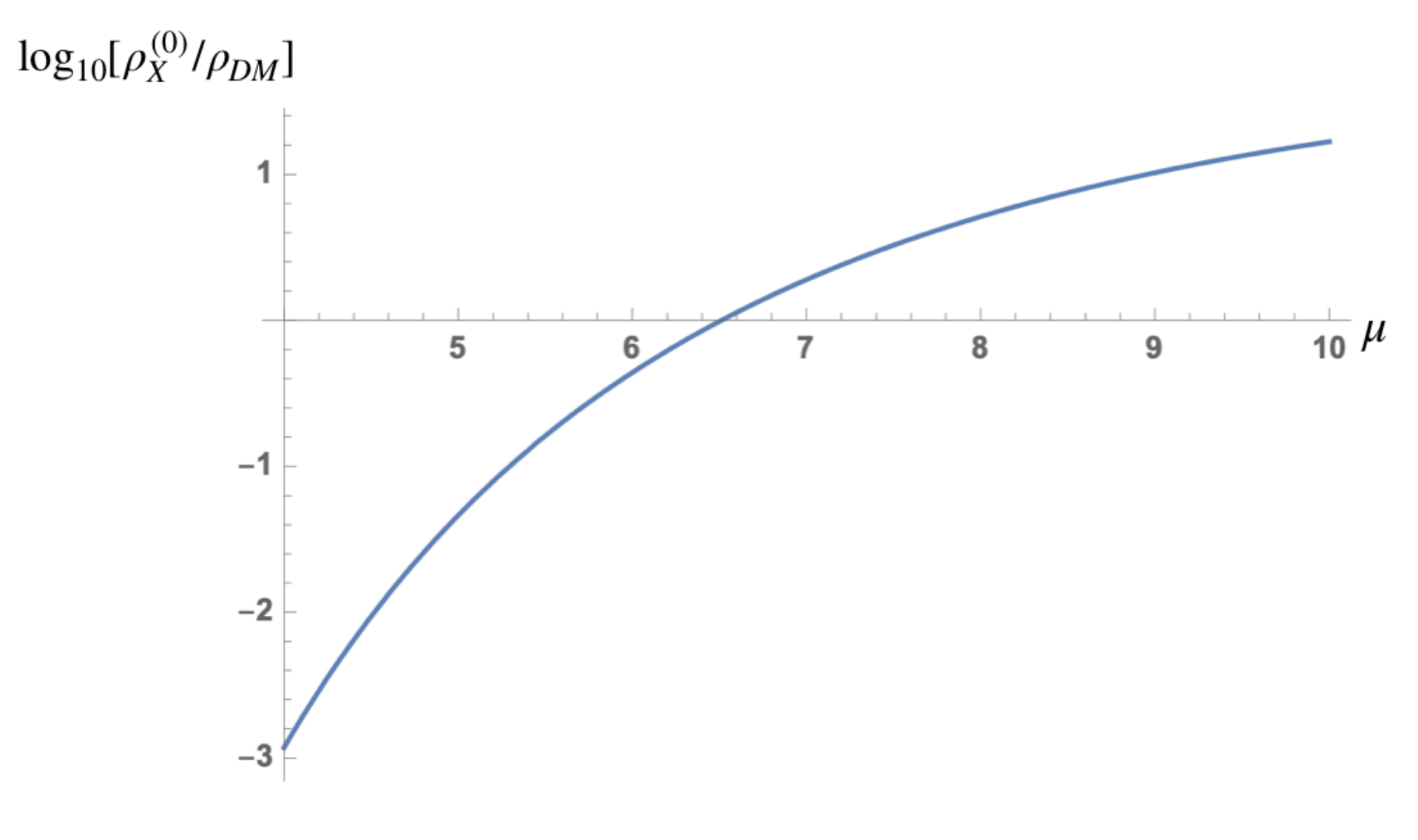}
 \caption{ Log of the ratio of the energy density of $X$-particles  to the  observed energy density of dark matter as a function 
  of $\mu = m_R/M_X$.   }
  \label{rho-of-MX}
 \end{figure}

According to our results, the mass of DM particles, with the interaction strength typical for supersymmetric ones, can
be in the range from $10^6$ to $10^{13} $ GeV. It is tempting to find if and how they could be observed. There are some possibilities to make $X$-particles visible. The first one may be connected with the annihilation effects in clusters of dark matter in galaxies and galactic halos,  in which, according to Refs.~\cite{Berezinsky:2014wya,Berezinsky:1996eg},  the density of DM is much higher than DM cosmological density. Another possibility is to consider 
the decay of superheavy DM particles, which could have a lifetime long enough to manifest themselves as stable dark matter, but at the same time lead to the  possibly observable contribution to the  UHECR spectrum. Furthermore, instability of superheavy DM particles can arise due to 
Zeldovich mechanism through virtual black holes formation~\cite{Zeldovich:1976vq,Zeldovich:1977be}. 

\section{Conclusion}

The existence of stable particles
with interaction strength typical for supersymmetry and heavier than several TeV is in tension with the conventional Friedmann cosmology. 
Starobinsky inflationary model opens a way to save life of such $X$-particles, because in this model the density of heavy relics  could be
significantly reduced. If the epoch of the domination of the curvature oscillations (the scalaron domination) lasted after freezing of massive species, their
density with respect to the plasma entropy could be noticeably diluted by  radiation from the scalaron decay. 

The range of allowed masses depends upon the dominant decay mode of  scalaron. If the scalaron is minimally
coupled to scalar particles $X_S$, the decay amplitude does not depend upon  the scalar particle mass and 
an acceptably low density of $X_S $ can be achieved if ${M_{X_S} \geq m_R \approx 3\cdot 10^{13}}$ GeV. 

If the scalaron predominantly decays into fermions or conformally coupled scalars then the probability of the  decay 
is proportional to the particle mass squared, and the allowed mass of $X$-particles to form DM could be ${m_X \sim 10^6}$ GeV.

If the scalaron decays into gauge bosons 
due to conformal anomaly and $X$-particles are Majorana fermions, they are thermally produced in plasma and
could make proper amount of dark matter, 
 if their mass is about $5 \cdot 10^{12}$ GeV. 

\section*{Acknowledgments}
This work was done under support by the RSF Grant 20-42-09010.


\end{document}